\documentclass[prl,twocolumn,showpacs,nofootinbib,aps,10pt,superscriptaddress,amsmath,amssymb]{revtex4-1}

\usepackage{graphicx}
\usepackage{dcolumn}
\usepackage{bm}
\usepackage{epsfig,color}
\usepackage{epstopdf}
\usepackage{mathtools}
\usepackage{natbib}
\usepackage{captcont}
\usepackage{booktabs}
\usepackage{threeparttable}
\usepackage{slashed}
\usepackage{extarrows}

\def\be{\begin{eqnarray}}
\def\en{\end{eqnarray}}
\def\non{\nonumber}

\def\B{{\cal B}}

\begin{document}

\title{The strangest lifetime: A bizarre story of $\tau(\Omega_c^0)$}

\author{Hai-Yang Cheng}
\email[e-mail: ]{phcheng@phys.sinica.edu.tw}
\affiliation{Institute of Physics, Academia Sinica, Taipei, Taiwan 11529, R.O.C.}

\maketitle

For ground-state singly charmed baryons, the $\Lambda_c^+$,
$\Xi_c^+$ and $\Xi_c^0$ form an antitriplet representation and
they all decay weakly. The $\Omega_c^0$, $\Xi'^+_c$, $\Xi'^0_c$
and $\Sigma_c^{++,+,0}$ form a sextet representation; among them,
only $\Omega_c^0$ decays weakly. Back in 1986 their lifetime pattern had already been predicted to be
\cite{Shifman:1986mx,Guberina:1986gd}
\be \label{eq:Bclifetimehierarchy}
\tau(\Xi_c^+)>\tau(\Lambda_c^+)>\tau(\Xi_c^0)>\tau(\Omega_c^0).
\en
The measured lifetime hierarchy appeared for the first time in the 1996 version of the Particle Data Group (PDG) after the lifetime of $\Omega_c^0$, the last piece of the four charmed baryons, was successfully measured in the fixed target experiments E687 \cite{E687:1995cvt} and WA89 \cite{WA89:1995lbz} in 1995. However, the early data had a wide spread. The situation was substantially improved  by the FOCUS experiment performed during the period of 2001-2003 \cite{FOCUS:Lambdac,FOCUS:Xicp,FOCUS:Xic0,FOCUS:Omegac}. According to the 2004 version of PDG \cite{PDG2004}, the world averages of their lifetimes then were given by
\begin{eqnarray} \label{eq:exptlifetime}
&& \tau(\Lambda^+_c)= (200\pm 6)\,{\rm fs}, \quad
\tau(\Xi^+_c)= (442\pm 26)\,{\rm fs},
  \nonumber \\
&& \tau(\Xi^0_c)= (112^{+13}_{-10})\,{\rm fs}, \quad~~
\tau(\Omega^0_c)= (69\pm12)\,{\rm fs}.
\end{eqnarray}
These world averages  remained stable from 2004 till 2018 \cite{PDG2018}. Notice that the charmed baryon lifetime pattern is quite different from the bottom baryon case where the lifetime hierarchy reads \cite{PDG}
\be
\tau(\Omega_b^-)>\tau(\Xi_b^-)>\tau(\Xi_b^0)\simeq\tau(\Lambda_b^0).
\en
That is, the $\Omega_b^-$ has the longest lifetime in the bottom baryon sector, contrary to the shortest lifetime of the $\Omega_c^0$ in the counterpart of charmed baryons.

Lifetimes of the heavy baryons are commonly analyzed within the framework of heavy quark expansion (HQE). In this general approach, the predicted lifetime hierarchy for charmed baryons given in Eq. (\ref{eq:Bclifetimehierarchy}) agrees with experimental pattern exhibited in Eq. (\ref{eq:exptlifetime}). The fact that the $\Omega_c$ is shortest-lived among the four charmed baryons owing to its large constructive Pauli interference (PI) has been known to the community for a long time.
However, the situation was dramatically changed in 2018 when LHCb reported a new measurement of the charmed baryon $\Omega_c^0$ lifetime using semileptonic $b$-hadron decays \cite{LHCb:tauOmegac}.  More precisely, LHCb found $\tau(\Omega_c^0)=(268\pm24\pm10\pm2)\,{\rm fs}$, using the semileptonic decay $\Omega_b^-\to\Omega_c^0\mu^-\bar \nu_\mu X$ followed by $\Omega_c^0\to pK^- K^-\pi^+$. This value is nearly four times larger than the 2018 world-average value of $\tau(\Omega_c^0)$ (see Eq. (\ref{eq:exptlifetime})) extracted from fixed target experiments. As a result, a new lifetime pattern emerged
\be \label{eq:newhierarchy}
\tau(\Xi_c^+)>{\tau(\Omega_c^0)}>\tau(\Lambda_c^+)>\tau(\Xi_c^0).
\en

The LHCb observation of a huge jump of the $\Omega_c^0$ baryon lifetime in 2018 is very  striking from both experimental and theoretical points of view.  This is the first time in the history of particle physics that the lifetime of a hadron measured in a new experiment was so drastically different from the old one. The LHCb  has collected 978 events of the $b$-tagged $\Omega_c^0$ decays which is about five times larger than those accumulated by all predecessors FOCUS, WA89 and E687 of fixed target experiments \cite{LHCb:tauOmegac}. As stressed in Ref. \cite{Bigi}, the lifetime value measured is so large that could have been easily measured much earlier than 2018 by experiments at $e^+e^-$ colliders whose resolution is about 150 fs typically. Since CLEO-c and Belle have both observed  the $\Omega_c^0$ and measured its mass, they should/could have measured quite easily the lifetime value as done by LHCb \cite{Bigi}.

\begin{figure}[b]
\begin{center}
\includegraphics[height=45mm]{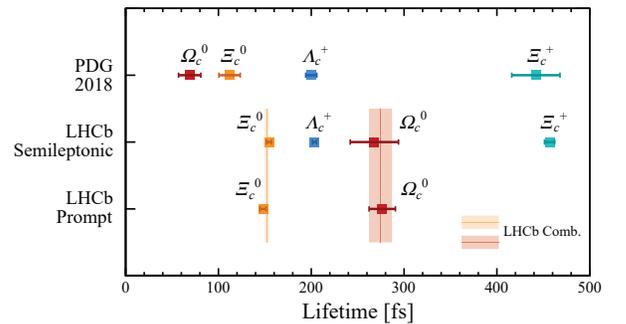}
\caption{Previous world-average values of the charmed baryon lifetimes from the PDG \cite{PDG2018} and the LHCb measurements of the $\Omega_c^0$ and $\Xi_c^0$ lifetimes obtained from semileptonic $\Omega_b^-$ decays and prompt signals. The combined LHCb results are shown in coloured bands. This figure is taken from the supplementary material provided in Ref. \cite{LHCb:tauOmegac_2}. }
\label{fig:charmbarylife}
\end{center}
\end{figure}

In 2019, LHCb has reported precision measurements of the $\Lambda_c^+$, $\Xi_c^+$ and $\Xi_c^0$ lifetimes \cite{LHCb:2019ldj} as displayed in Table \ref{tab:expt_lifetimes} and Fig. \ref{fig:charmbarylife}. The $\Xi_c^0$ baryon lifetime is approximately 3.3 standard deviations larger than the world average value.
Finally, this year LHCb \cite{LHCb:tauOmegac_2} reported a new measurement using promptly produced $\Omega_c^0$ and $\Xi_c^0$ baryons with 5.4 fb$^{-1}$ of the LHCb data in which $\Omega_c^0$ and $\Xi_c^0$ were reconstructed through their decays to $pK^-K^+\pi^+$. The results are
\be
\tau(\Omega_c^0) &=& (276.5\pm13.4\pm4.4\pm0.7)\,{\rm fs}, \non \\
\tau(\Xi_c^0) &=& (148.0\pm2.3\pm2.2\pm0.2)\,{\rm fs}.
\en
Hence, the previous LHCb measurement of $\tau(\Omega_c^0)$ based on the semileptonic decays of $\Omega_b^-$ is confirmed and its precision is improved by a factor of 2. As a consequence, the new lifetime hierarchy Eq. (\ref{eq:newhierarchy}) is now firmly established.

\begin{widetext}
\begin{table*}[t]
\caption{Evolution of the charmed baryon lifetimes measured in units of ${\rm fs}$.}
\label{tab:expt_lifetimes}
\begin{center}
\begin{tabular}{l c c c c} \hline \hline
 & $\tau(\Xi_c^+)$ & $\tau(\Lambda_c^+)$ & $\tau(\Xi_c^0)$ & $\tau(\Omega_c^0)$ \\
\hline
PDG (2004-2018) \cite{PDG2018}~~~ & ~$442\pm 26$ & $200\pm6$ & $112^{+13}_{-10}$ & $69\pm12$\\
LHCb (2018) \cite{LHCb:tauOmegac} & & & & ~~$268\pm26$~~ \\
LHCb (2019) \cite{LHCb:2019ldj} & $457\pm6$ & $203.5\pm2.2$ & $154.5\pm2.6$ & \\
PDG (2020) \cite{PDG} & ~~$456\pm5$~~ & ~~$202.4\pm3.1$~~ & ~~$153\pm 6$~~ & ~~$268\pm26$~~ \\
LHCb (2021) \cite{LHCb:tauOmegac_2} & & & $148.0\pm3.2$ & $276.5\pm14.1$ \\
World average (2021) & ~~$456\pm 5$~~ & ~~$202.4\pm3.1$~~ & ~~$152.0\pm 2.0$~~ & ~~$274.5\pm12.4$~~ \\
\hline \hline
\end{tabular}
\end{center}
\end{table*}
\end{widetext}

\begin{figure}[t]
\begin{center}
\includegraphics[height=14mm]{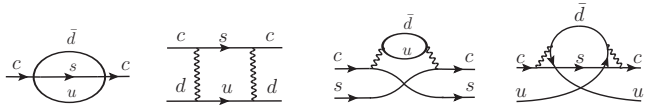}
\caption{Contributions to nonleptonic decay rates of charmed
baryons from four-quark operators (from left to right): $c$-quark decay,
$W$-exchange, constructive PI and
destructive PI. Note that the constructive PI is unique to the charmed baryon sector, as it does not occur in the bottom sector.
} \label{fig:fourquarkNL}
\end{center}
\end{figure}

On the theoretical side, HQE in powers of $1/m_Q$ is the standard theoretical framework for analyzing the lifetimes of bottom and charmed hadrons.
On the basis of the operator product expansion (OPE), the inclusive rate of the charmed baryon $\B_c$ is schematically represented by
 \begin{eqnarray} \label{eq:OPE}
 \Gamma({\cal B}_c\to f) = {G_F^2m_c^5\over
192\pi^3}V_{\rm CKM}\left(A_0+{A_2\over m_c^2}+{A_3\over
m_c^3}+\cdots\right)
 \end{eqnarray}
where $V_{\rm CKM}$ is the relevant Cabibbo-Kobayahsi-Maskawa matrix element.
The $A_0$ term arises from the decay of the heavy $c$ quark. There are no linear $1/m_c$ corrections known as Luke's theorem.  Nonperturbative corrections start at order $1/m_c^2$ which come from the interaction of the heavy quark spin with the gluon field .
The $A_3$ term consists of the four-quark operators $(\bar Q\bar\Gamma q)(\bar q\Gamma Q)$ with $\Gamma$ representing a combination of the Lorentz and color matrices. Spectator effects induced by dimension-6 four-quark operators include $W$-exchange, constructive and destructive  PI as depicted in
Fig. \ref{fig:fourquarkNL}. Although spectator effects are formally of
order $1/m_c^3$, they gain a phase-space enhancement factor of $16\pi^2$ relative to the three-body phase space for heavy quark decay. Consequently, spectator effects
are comparable to and even exceed the $1/m_c^2$ terms. HQE to the order of $1/m_b^3$ works very well for $B$ mesons and bottom baryons (see e.g. Ref. \cite{Cheng:2018}).

The total width of the charmed baryon ${\cal B}_c$
receives contributions from inclusive nonleptonic (NL) and semileptonic (SL) decays.
The NL contribution can be further decomposed into
 \begin{eqnarray}
 \Gamma_{\rm NL}({\cal B}_c)=\Gamma^{\rm dec}({\cal B}_c)+\Gamma^{\rm ann}
 ({\cal B}_c)+\Gamma^{\rm
 int}_+({\cal B}_c)+\Gamma^{\rm int}_-({\cal B}_c), \non \\
 \end{eqnarray}
corresponding to the $c$-quark decay, $W$-exchange
contribution, constructive and destructive PI, respectively.
For inclusive SL decays, there is an additional spectator effect in charmed-baryon
SL decay originating from the PI of the $s$
quark \cite{Voloshin} similar to the constructive PI in Fig. \ref{fig:fourquarkNL} with $\bar d$ and $u$ quarks replaced by $\ell^+$ and $\nu_\ell$, respectively.

Because the $\Omega_c^0$ contains two $s$ quarks, it is natural to expect that
\be
&& \Gamma^{\rm
int}_+(\Omega_c^0)\gg \Gamma^{\rm int}_+(\Xi_c^{+,0})\gg \Gamma^{\rm int}_+(\Lambda_c^+), \non \\
&& \Gamma^{\rm
SL}(\Omega_c^0)\gg \Gamma^{\rm SL}(\Xi_c^{+,0})\gg \Gamma^{\rm SL}(\Lambda_c^+).
\en
That is, the  $\Omega_c^0$ is expected to be shortest-lived as it receives largest constructive PI in both NL and SL decays.
It turns out that the observed old lifetime hierarchy $\tau(\Xi_c^+)>\tau(\Lambda_c^+)>\tau(\Xi_c^0)>\tau(\Omega_c^0)$ which had been settled for a long time until 2018 can be understood at least qualitatively  in the OPE approach up to $1/m_c^3$ expansion.
Therefore, the LHCb observation of a much longer $\tau(\Omega_c^0)$ is beyond imagination. Does it mean that HQE fails to apply to the charmed baryon system?

It should be stressed that although the qualitative feature of the lifetime pattern is comprehensible, the quantitative estimates of charmed baryon lifetimes and their ratios are
still rather poor. For example, $R_1\equiv\tau(\Xi_c^+)/\tau(\Lambda_c^+)$ and $R_2\equiv \tau(\Xi_c^+)/\tau(\Xi_c^0)$ are calculated to be, for example, 1.03 and 1.90, respectively, in Ref. \cite{Cheng:2018} while experimentally $R_1=2.21\pm0.15$,  $R_2=3.95\pm0.47$ from 2018 PDG \cite{PDG2018} and  $2.30\pm0.04$,  $3.00\pm0.05$ from 2021 world average. Hence,  contrary to $B$ meson and bottom baryon cases where HQE in $1/m_b$ leads to the lifetime ratios in excellent agreement with experimental results, HQE in $1/m_c$ does not work well for describing the lifetime ratios of charmed baryons. Since the charm quark is not heavy enough, it is sensible to consider the subleading $1/m_c$ corrections to spectator effects. This means that we need to consider $1/m_c^4$ expansion in Eq. (\ref{eq:OPE}) with dimension-7 operators which are either the four-quark operators times the spectator quark mass or the four-quark operators with an additional derivative \cite{Gabbiani}.

An analysis of  the subleading $1/m_c$ corrections to spectator effects described by dimension-7 operators in  Ref. \cite{Cheng:2018} indicated that while the results on $\Lambda_c^+$ and $\Xi_c^+$ were improved (see Table \ref{tab:lifetime:final}), HQE failed to apply to the $\Omega_c^0$. To see the issue of HQE with the $\Omega_c^0$, we write
$\Gamma_+^{\rm int}(\Omega_c)=\Gamma^{\rm int}_{+,6}(\Omega_c)+\Gamma^{\rm int}_{+,7}(\Omega_c)$ and  $\Gamma^{\rm SL}(\Omega_c)=\Gamma^{\rm SL}_{6}(\Omega_c)+\Gamma^{\rm SL}_{7}(\Omega_c)$. It turns out that the dimension-7 contributions $\Gamma^{\rm int}_{+,7}(\Omega_c)$ and $\Gamma^{\rm SL}_7(\Omega_c)$ are destructive and their sizes are so large that they overcome the dimension-6 ones and flip the sign (see Table X of Ref. \cite{Cheng:2021qpd}). Of course, a negative $\Gamma^{\rm SL}(\Omega_c)$ does not make sense and the subleading corrections are too large to justify the validity of HQE. Hence, HQE fails to apply to the $\Omega_c^0$ to order $1/m_c^4$!
Nevertheless, by demanding a sensible HQE for the $\Omega_c^0$, we might introduce an {\it ad hoc} parameter $\alpha$ to suppress the contributions from dimension-7 operators to render both $\Gamma^{\rm SL}(\Omega_c)$ and $\Gamma^{\rm int}_{+}(\Omega_c)$ positive. It was  conjectured in Ref. \cite{Cheng:2018} that $\alpha$ lies in $0.16<\alpha<0.32$ and the $\Omega_c^0$ lifetime falls in the range of $230\sim330\,{\rm fs}$. This leads to a lifetime of the $\Omega_c^0$ longer than $\Lambda_c^+$ and hence the new hierarchy Eq. (\ref{eq:newhierarchy}).
We conclude that the $\Omega_c^0$, which is naively expected to be shortest-lived in the charmed baryon system owing to the large constructive PI, could live longer than the $\Lambda_c^+$  due to the suppression from $1/m_c$ corrections arising from dimension-7 four-quark operators.

\begin{table}[t]
\caption{Calculated lifetimes  of charmed baryons (in units of fs) in the HQE to order $1/m_c^3$ and $1/m_c^4$ including the suppression parameter $\alpha$ for the $\Omega_c^0$ \cite{Cheng:2018}. New world averages (WA) are taken from Table \ref{tab:expt_lifetimes}.
}
\label{tab:lifetime:final}
\begin{center}
\begin{tabular}{c c c c } \hline \hline
 & ~~HQE ($1/m_c^3$) & ~HQE ($1/m_c^4$ with $\alpha$)~ & 2021 WA \\
\hline
 $\Xi_c^+$ & 306 & 392 & $456\pm5$  \\
 $\Omega_c^0$ & 103 & 230$\sim$330  & $274.5\pm12.4$    \\
 $\Lambda_c^+$ & 296 & 212 & $202.4\pm3.1$ \\
 $\Xi_c^0$ & 161 & 156 & $152.0\pm2.0$ \\
\hline \hline
\end{tabular}
\end{center}
\end{table}

The calculated charmed baryon lifetimes in HQE to order $1/m_c^3$ and $1/m_c^4$ (including $\alpha$ for the $\Omega_c^0$) are summarized in Table \ref{tab:lifetime:final}. The $\Omega_c^0$ lifetime cannot be predicted precisely, as it depends on the unknown parameter $\alpha$.
The origin of this suppression parameter is unknown, but it could be due to the next-order $1/m_c$ correction. It has been three years since the first new measurement of $\tau(\Omega_c^0)$ by LHCb.  In spite of the progresses in regard to the lifetimes of heavy mesons \cite{Kirk:2017juj,King:2021xqp},
theoretical activity concerning the $\Omega_c^0$ lifetime is still absent.
It is hoped that this bizarre story of the $\Omega_c^0$ lifetime could get QCD theorists thinking seriously about how to refine the relevant calculations and work out the higher order terms in HQE.

{\it Acknowledgments}

This research was supported in part by the Ministry of Science and Technology of R.O.C. under Grant No. MOST-110-2112-M-001-025.


\end{document}